\documentclass[aps,prd,superscriptaddress,twocolumn,nofootinbib]{revtex4-1}

\usepackage[colorlinks=true,citecolor=green,urlcolor=blue]{hyperref}
\usepackage{amssymb}
\usepackage{amsmath,color}
\usepackage{graphicx}
\usepackage{bbm}
\usepackage{subfigure}
\usepackage{verbatim}
\usepackage[T1]{fontenc}
\usepackage[utf8]{inputenc}
\usepackage[normalem]{ulem}
\usepackage{longtable}

\usepackage{url}
\usepackage{xcolor}

\begin{document}

\title{Model-Independent Perspectives on Coupled Dark Energy and the Swampland}
\author{Tao Yang}
\email{tao.yang@apctp.org}
\affiliation{Asia Pacific Center for Theoretical Physics, Pohang 37673, Korea}

\date{\today}

\begin{abstract}
We present a general model-independent approach to study the coupled dark energy and the string Swampland criteria. We show how the dark sector interaction is degenerated with the equation of state of dark energy in the context of the expansion of the Universe. With priors for either of them, the dynamics of dark energy and the dark sector interactions can be reconstructed together with the bounds of the Swampland criteria. Combining cosmic chronometers, baryon acoustic oscillation (BAO), and Type Ia supernovae our results suggest a mild $1 \sigma$ significance of dark sector interactions at low redshift for the coupled quintessence. The Lyman-$\alpha$ BAO at $z=2.34$ leads a $2 \sigma$ signal of nonzero interactions at high redshift. The implications for coupled quintessence are discussed.

\end{abstract}

\maketitle

\section{Introduction} \label{sec:intro}

The concordance cosmology--a cosmological constant $\Lambda$ plus the cold dark matter (CDM), namely the $\Lambda$CDM model, has proven to be remarkably successful at describing the Universe from a wide range of experiments including cosmic microwave background (CMB), baryon acoustic oscillations (BAO), Type Ia supernovae (SNe Ia), and large-scale structures (LSS)~\citep{Riess:1998cb,Perlmutter:1998np,Hinshaw:2012aka,Aghanim:2018eyx,Alam:2016hwk}. Despite the consistency with the observations, $\Lambda$CDM model still faces several problems~\citep{Weinberg:2000yb}. In addition, tensions between different experiments when fitting the $\Lambda$CDM model have emerged in the last decade (see the Hubble tensions~\citep{Freedman:2017yms,Verde:2019ivm,Aghanim:2018eyx,Wong:2019kwg,Riess:2019cxk} and cosmic shear discrepancies~\citep{McCarthy:2017csu,Hildebrandt:2018yau,Asgari:2019fkq}). These discordances, if not induced by unknown systematics, challenge the standard cosmological model and hint at a fundamental problem or new physics. Many attempts, such as introducing dynamic dark energy (see reviews~\citep{Copeland:2006wr,Tsujikawa:2013fta}) or more fundamentally modify the general relativity (GR)~\citep{Kunz:2006ca,Tsujikawa:2010zza,Ade:2015rim}, have been proposed on these issues. 
Among these extensions to the baseline $\Lambda$CDM model, the interaction between dark energy and dark matter have also been studied in the literature~\citep{Wang:2016lxa,Amendola:1999er,Das:2005yj,Chimento:2013rya,Richarte:2014yva,Li:2014cee,Costa:2016tpb,DAmico:2016jbm,vandeBruck:2017idm,Zhang:2018mlj,Li:2018ydj,DAmico:2018mnx}, which is aimed at solving the coincidence problem and may be an excellent solution to the $H_0$ and cosmic shear tensions~\citep{Mangano:2002gg,Zhang:2005rg,Yang:2018euj,DiValentino:2019jae,DiValentino:2019ffd}.

Quintessence as a canonical scalar field is introduced to explain cosmic acceleration with a dynamical cosmological constant~\citep{Tsujikawa:2013fta,Linder:2007wa}. It is also argued that unless there is a symmetry, a quintessence field should couple to other sectors~\citep{Carroll:1998zi,Amendola:1999er,Zhang:2005rj,Han:2018yrk}. Given the tight constraint of the couplings to the standard model particles, the scaler field is at least considered coupling to dark matter, namely the ``coupled quintessence''~\citep{Amendola:1999er,Das:2005yj}. Coupled dark energy is closely related to modified theories of gravity, for example, there is a 
conformal equivalence between the $f(R)$ gravity in the Jordan frame and the coupled dark energy model in the Einstein frame~\citep{DeFelice:2010aj,Wang:2016lxa}. These facts make the interactions between dark energy and dark matter a general consideration that cannot be excluded a priori.

Either the coupled phenomenological fluid or the quintessence of dark energy models have been constrained by various data sets in the literature~\citep{Li:2014cee,Costa:2016tpb,vandeBruck:2017idm,Zhang:2018mlj,An:2018vzw,Mifsud:2017fsy}. Usually, a specific form of the dark energy model together with a coupling should be assumed to obtain the constraints from the model-fitting methodology (see~\citep{vandeBruck:2017idm} for exponential coupling and exponential potential of scalar field). A nonparametric and model-independent approach was proposed by~\citet{Cai:2015zoa} to 
reconstruct the interaction between the phenomenological fluid of dark energy and dark matter. In this paper, we investigate the coupled dark energy for both the phenomenological fluid and quintessence in the context of the model-independent approach.

Recently, the string Swampland criteria on cosmology has been proposed~\citep{Agrawal:2018own,Obied:2018sgi,Vafa:2005ui,Garg:2018reu,Ooguri:2018wrx}.
A common viewpoint is that GR cannot be the ultimate theory of the Universe. We can however assume that GR might be the low-energy limit of a well-motivated high-energy UV-complete theory such as the string theory.  Thus, effective field theory (EFT) describes these low-energy limits that effectively capture the behavior of the inflaton field and dark energy phenomena. The landscape of string theory gives a vast range of choices for how our Universe may fit in a consistent quantum theory of gravity. However it is surrounded by an even bigger swampland of consistent-looking semiclassical EFTs, which are actually inconsistent~\citep{Vafa:2005ui}. This could be an indicator that de Sitter vacua may reside in the Swampland. 

One of the Swampland criteria~\footnote{Another Swampland criterion is the distance conjecture which says, {\it the range traversed by scalar fields in field space is bounded by $\frac{\Delta \phi}{M_{\rm pl}}\lesssim\mathcal{O}(1)$.}} comes from the refined de Sitter conjecture: {\it Any scalar field potential from string theory obeys either $M_{\rm pl}\frac{V_{\prime\phi}}{V}\equiv c\gtrsim\mathcal{O}(1)$ (C1)
or $M_{\rm pl}^2\frac{V_{\prime\prime\phi}}{V}\equiv \tilde c\lesssim -\mathcal{O}(1)$ (C2)}~\citep{Garg:2018reu,Ooguri:2018wrx}.
Here $\phi$ is the scalar field and $V$ is the potential, which are usually built form the quintessence model. Throughout this paper, we set the reduced Planck mass to be $M_{\rm pl}=1$.
The Swampland criteria have sparked a lot of research recently on cosmology~\citep{Heisenberg:2018yae,Raveri:2018ddi,Cai:2018ebs,Brahma:2019kch,vandeBruck:2019vzd,Heisenberg:2020ywd},
especially on $H_0$ tensions~\citep{Colgain:2018wgk,Colgain:2019joh,Anchordoqui:2019amx,Banerjee:2020xcn}.
We know usually the quintessence models exacerbate the Hubble tensions because of the negative correlation between $H_0$ and $w$ (see examples in~\citep{DiValentino:2017zyq,Yang:2018uae} and references therein). Recently,~\citet{Agrawal:2019dlm} proposed an interacting quintessence model inspired from distance conjecture, which leads to a continually reducing dark matter mass as the scalar field rolls in the recent cosmological epoch, can reconcile the Hubble tension. However, the couplings emerge from $z\approx15$ where the observations are rare.

Like the dark sector interaction, traditional constraints of the Swampland criteria confronting the cosmological data sets were performed by the model-fitting methodology~\citep{Heisenberg:2018yae,Akrami:2018ylq}. In particular,~\citet{Wang:2018duq} fitted the cosmological model with a large parameter space including the coupling to various data sets . Recently, a model-independent approach was proposed by~\citep{Elizalde:2018dvw} to reconstruct $C1$ but they assumed no interaction in the dark sector, which is not general in the literature.

In this paper, without assuming any cosmological models, we investigate a model-independent approach to reconstruct the coupled dark energy and the bounds of the Swampland criteria. We would like to present a most general model-independent data analysis in the context of the expansion of the Universe to study the coupled dark energy and string theory implications on cosmology.

\section{Method setup} \label{sec:method}

We consider a quintessence scalar field $\phi$ coupled to the non-relativistic dark matter.  In the presence of this dark-sector interaction, the dark matter density no longer scales as $a^{-3}$ but $\rho_c\sim f/a^3$.
Here $f$ is an arbitrary function of $\phi$ to denote the coupling~\citep{Das:2005yj}.
To be consistent with the equivalence principle and Solar System tests of gravity~\citep{Bertotti:2003rm,Will:2014bqa}, we do not
couple $\phi$ to baryons~\footnote{Actually, introducing a universal coupling between dark energy and the total non-relativistic matter including the subdominant baryons only slightly changes our results.}. Without loss of generality, we assume at present time $f_0=1$.
In the flat Friedmann-Lema\^\i tre-Robertson-Walker (FLRW) spacetime, the Friedmann and continuity equations for every species are
\begin{eqnarray}
&3H^2=\rho_\phi+\rho_c+\rho_b\,,\\
&\dot\rho_c+3H\rho_c=3H_0^2\Omega_c\frac{\dot f}{a^3} \,,\\
&\dot\rho_{\phi}+3H(1+w_{\phi})\rho_{\phi}=-3H_0^2\Omega_c\frac{\dot f}{a^3} \,, \\
&\dot\rho_b+3H\rho_b=0 \,.
\end{eqnarray}
Here $a$ is the scale factor, $H=\dot a/a$ is the Hubble parameter and dots denote time derivatives.
The energy density and pressure of the scalar field are
$\rho_{\phi}=\frac{1}{2}\dot \phi^2+V$ and $P_{\phi}=\frac{1}{2}\dot \phi^2-V$, with the field potential $V$ and equation of state $w_\phi=P_\phi/\rho_\phi$.
$\Omega_X=\rho_{X0}/3H_0^2$ denotes the present values of the density parameters.
Since we focus on the late-time Universe, the radiation contributions are ignored. 

After some algebra, we can get the equations for the kinetic term and potential of the scalar field,
\begin{eqnarray}
\dot \phi^2&=&-2\dot H-3H_0^2\frac{\Omega_cf+\Omega_b}{a^3} \,, \label{eq:dotphi2}\\
V&=&\dot H+3H^2 -\frac{3}{2}H_0^2\frac{\Omega_cf+\Omega_b}{a^3} \,, \label{eq:V}\\
\dot V&=&\ddot H+6H\dot H -\frac{3}{2}H_0^2\frac{\Omega_c\dot f-3H(\Omega_cf+\Omega_b)}{a^3} \,. \label{eq:dotV}
\end{eqnarray}
The above equations make it possible to relate the Swampland criteria to the expansion data sets of the Universe and the coupling between dark energy and dark matter. For the uncoupled case $f=1$, we recover the formulae in~\citep{Elizalde:2018dvw}.

To relate the coupling and the equation of state of the scaler field, 
$w_\phi=(-2\dot H-3H^2)/(3H^2-3H_0^2\frac{\Omega_cf+\Omega_b}{a^3})$. 
We have
\begin{eqnarray}
f=\frac{a^3}{3H_0^2\Omega_c} && \left(\frac{2\dot H+3H^2(1+w_{\phi})}{w_{\phi}}\right)-\frac{\Omega_b}{\Omega_c} \,, \label{eq:f}\\
\dot f=\frac{a^3}{3H_0^2\Omega_c} && \left((9H^3+6H\dot H)+ \frac{12H\dot H+2\ddot H+9H^3}{w_{\phi}} \right .\nonumber \\
&& \quad\left .-\frac{(2\dot H+3H^2)\dot w_{\phi}}{w_\phi^2}\right) \,. 
\label{eq:dotf}
\end{eqnarray}

 
Eqs.~(\ref{eq:f}) and (\ref{eq:dotf}) are also valid for the phenomenological dark energy with a general equation of state. The phenomenological dark energy here is a general fluid with no prior restriction on the time-dependent equation of state $w(z)$.  The equation of state of dark energy is degenerated with the dark sector interactions in the context of the background expansion observations ({\it e.g.} Hubble parameter, distance). That is, for a fixed background expansion, $w_\phi$ (or a general $w$) and $f$ can be reconstructed only if either of them is given. In other words, the dynamics of the dark energy and dark sector interactions counterbalance each other to give the expansion of the Universe measured from the observations. Note that in order to reconstruct the functions from the data sets, we should convert the time derivatives to redshift derivatives in all equations through the relation $\frac{d}{dt}=-H(1+z)\frac{d}{dz}$. 

For the coupled quintessence, we can directly relate $\dot\phi^2$, $V$, $\dot V$ to the equation of state of the scaler field
\begin{eqnarray}
\dot \phi^2&=&-\frac{1+w_{\phi}}{w_{\phi}}(2\dot H+3H^2) \,, \label{eq:dotphi2w} \\
V&=&\frac{(w_{\phi}-1)(2\dot H+3H^2)}{2w_{\phi}} \,, \label{eq:Vw} \\
\dot V&=&\frac{w_\phi(w_\phi-1)(2\ddot H+6H\dot H)+\dot w_\phi(2\dot H+3H^2)}{2w_\phi^2} \,. \label{eq:dotVw}
\end{eqnarray}
The derivations of $\ddot\phi$ and $\ddot V$ are straightforward. Having the prior of $w_\phi$, we can reconstruct the dark sector interactions ($f$, $\dot f$) and the Swampland criteria $\frac{|V_{\prime\phi}|}{V}=\frac{|\dot V/\dot\phi|}{V}$ or $\frac{V_{\prime\prime\phi}}{V}=(\frac{\ddot V}{\dot\phi^2}-\frac{\ddot\phi \dot V}{\dot\phi^3})/V$ simultaneously.

Equations~(\ref{eq:dotphi2}-\ref{eq:dotVw}) suggest that using the the expansion data sets of the Universe, we can reconstruct the dynamics of the dark energy, the kinetic term and potential of the scalar field, the dark sector interactions, and the final swampland criteria if we set a prior for either the coupling $f$ or the equation of state of dark energy $w$ (for quintessence, $w_\phi>-1$ should be hold).

Instead of the Hubble parameter, the distance can also be incorporated in the equations above using the relations $H^{-1}\sim d(D_L/(1+z))/dz\sim d(D_A(1+z))/dz$. Here $D_L$ is the luminosity distance and $D_A$ is the angular diameter distance. In this paper, we stick to using the the Hubble observations.

We consider three data sets related to the expansion of the Universe:
\begin{itemize} 
\item 31 $H(z)$ data from cosmic chronometers (CC) obtained using the differential-age technique. We use the compilation in Table 1 of~\citet{Gomez-Valent:2018hwc} (also see references therein); 
\item 18 $H(z)$ data from the homogenized model-independent BAO compiled in Table 2 of~\citet{Magana:2017nfs}. But we substitute the last three high redshift $z=2.33,~2.34,~2.36$ Lyman-$\alpha$ (Ly$\alpha$) BAO~\citep{Font-Ribera:2013wce,Delubac:2014aqe,Bautista:2017zgn} with the latest update at $z=2.34$ from eBOSS DR14~\citep{Agathe:2019vsu,Blomqvist:2019rah}. Note we adopt the Planck sound horizon around 147 Mpc at the drag epoch;
\item 6 $E(z)$ data from Pantheon+MCT SNe Ia Measurements given by~\citet{Riess:2017lxs}. Here $E(z)=H(z)/H_0$ is the Hubble rate. For the $z=1.5$ point we adopt the Gaussian approximation by~\citet{Gomez-Valent:2018gvm}.
\end{itemize}

For the compatibility of the three data sets that we use for the reconstructions, we convert all the $H(z)$ data to $E(z)$ data by setting a fiducial value of $H_0$. Again, in the spirit of a model-independent approach, we set the fiducial value of $H_0$ estimated from the non-parametric reconstruction using CC+BAO $H(z)$ data. We get the mean $H_0=67.72~\rm km~s^{-1}~Mpc^{-1}$, which is consistent with Planck~\footnote{The lower value of $H_0$ is caused by the Ly$\alpha$ BAO at high redshift. We find without Ly$\alpha$ the reconstructed $H_0$ is $69.03~\rm km~s^{-1}~Mpc^{-1}$. However our approach is not sensitive to the value of $H_0$.}. We rewrite Eqs.~(\ref{eq:dotphi2}-\ref{eq:dotVw}) to substitute $H$ with $E$ by dividing $(H_0)^n$ on both sides of the equations. Here n relates to the dimensions of the equations quantified by $H_0$. Then the dimensionless parameters we can reconstruct are actually $\frac{\dot \phi^2}{H_0^2}$, $\frac{V}{H_0^2}$, $\frac{\dot V}{H_0^3}$, $f$, $\frac{\dot f}{H_0}$, $w_\phi$ (or $w$), $c$, and $\tilde c$, etc.

\section{Reconstructions} \label{sec:rec}

\begin{figure*}
\includegraphics[width=0.45\textwidth]{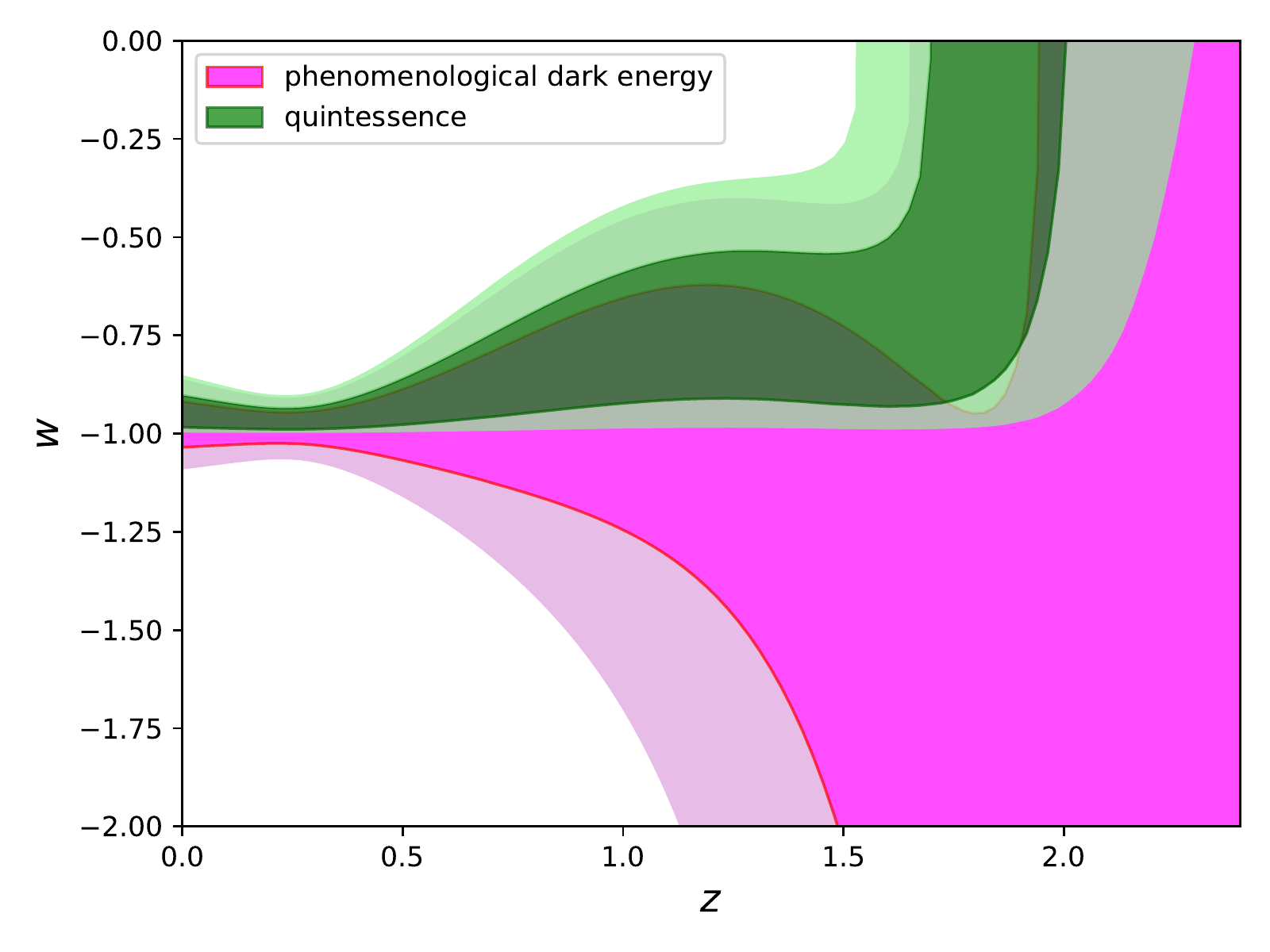} 
\includegraphics[width=0.45\textwidth]{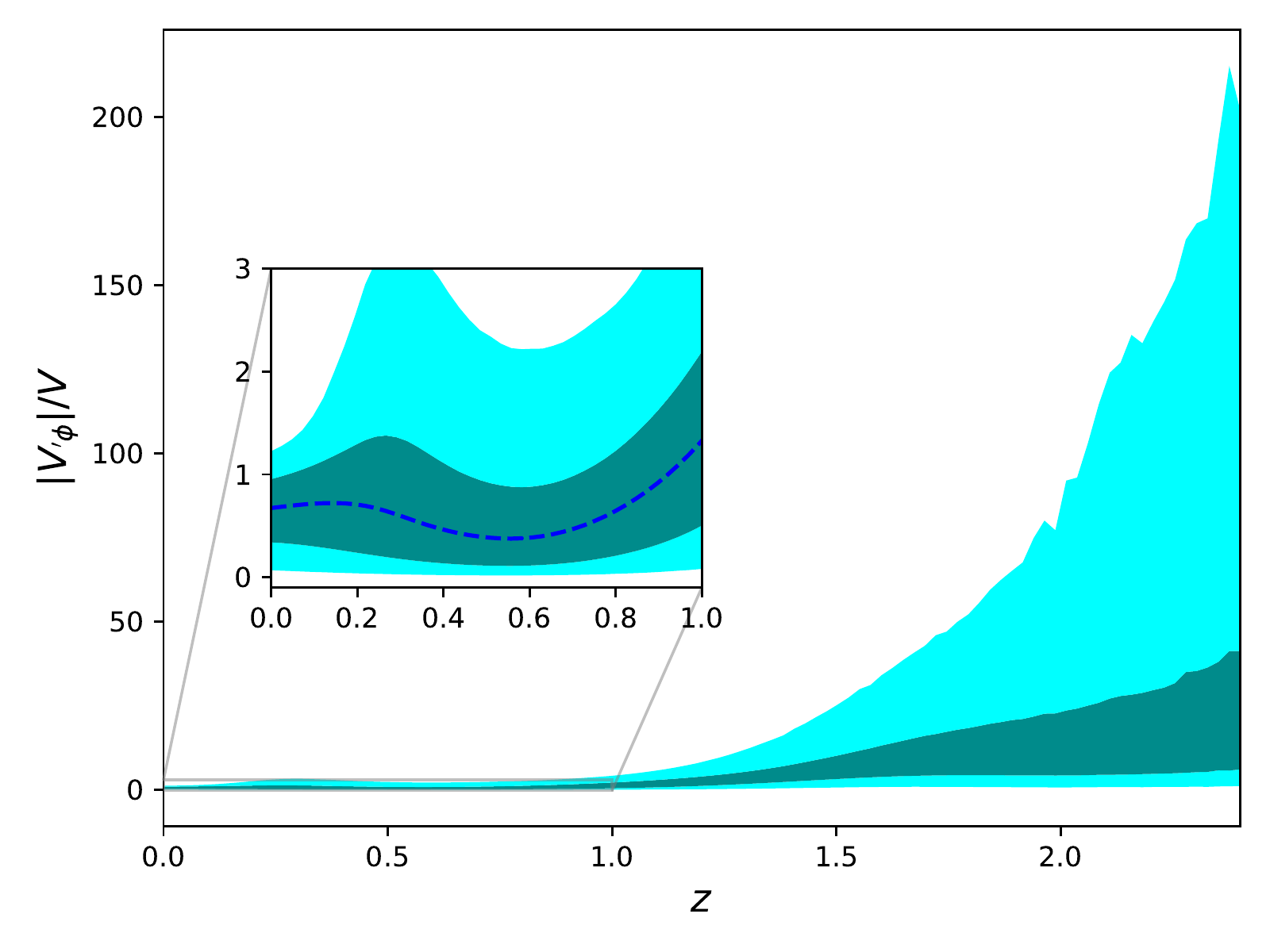} 
\caption{Reconstructions of the equation of state of dark energy (left) and Swampland criterion $C1$ (right) for the uncoupled dark energy. We show $68\%$ and $95\%$ confidence level with the dark and light colors. The dashed denotes the mean of the reconstruction.}
\label{fig:uncoupled}
\end{figure*}

We use Gaussian process~\footnote{\url{http://www.gaussianprocess.org/gpml/chapters/RW.pdf}}, a machine learning method which has been widely used in the data analysis of cosmology especially for the model-independent reconstructions of the cosmological parameters (see~\citep{Shafieloo:2012ht,Seikel:2012uu} and references therein), to reconstruct $E(z)$, $E'(z)$, $E''(z)$ and $E'''(z)$ from the three date sets combination. In this paper, we adopt the value $\Omega_c=0.26$ and $\Omega_b=0.05$ from Planck~\cite{Aghanim:2018eyx}.  From these $E^{(n)}(z)$ reconstructions with their covariance, we first check the uncoupled case in Eqs.~(\ref{eq:dotphi2}-\ref{eq:dotV}). For uncoupled quintessence, our reconstructed Swampland criteria $C1$ (see Fig.~\ref{fig:uncoupled}) is consistent with~\citet{Elizalde:2018dvw}. Note that for quintessence as dark energy, we should stick to  $w>-1$ (no phantom). From Eq.~(\ref{eq:dotphi2w}), this amounts to ensuring a positive value of the kinetic term $\dot\phi^2$. The equation of state of either the phenomenological dark energy or the quintessence is also shown in Fig.~\ref{fig:uncoupled}. The results show that the data sets
are consistent with the $\Lambda$CDM model ($w=-1$ and $\dot f=0$). The upper bound of the Swampland criterion $C1$ at $z=0$ is 1.23 at $95\%$ significance~\footnote{We do not show the results of C2 in this paper because of the poor reconstructions using present data sets.}.

\begin{figure*}
\includegraphics[width=0.45\textwidth]{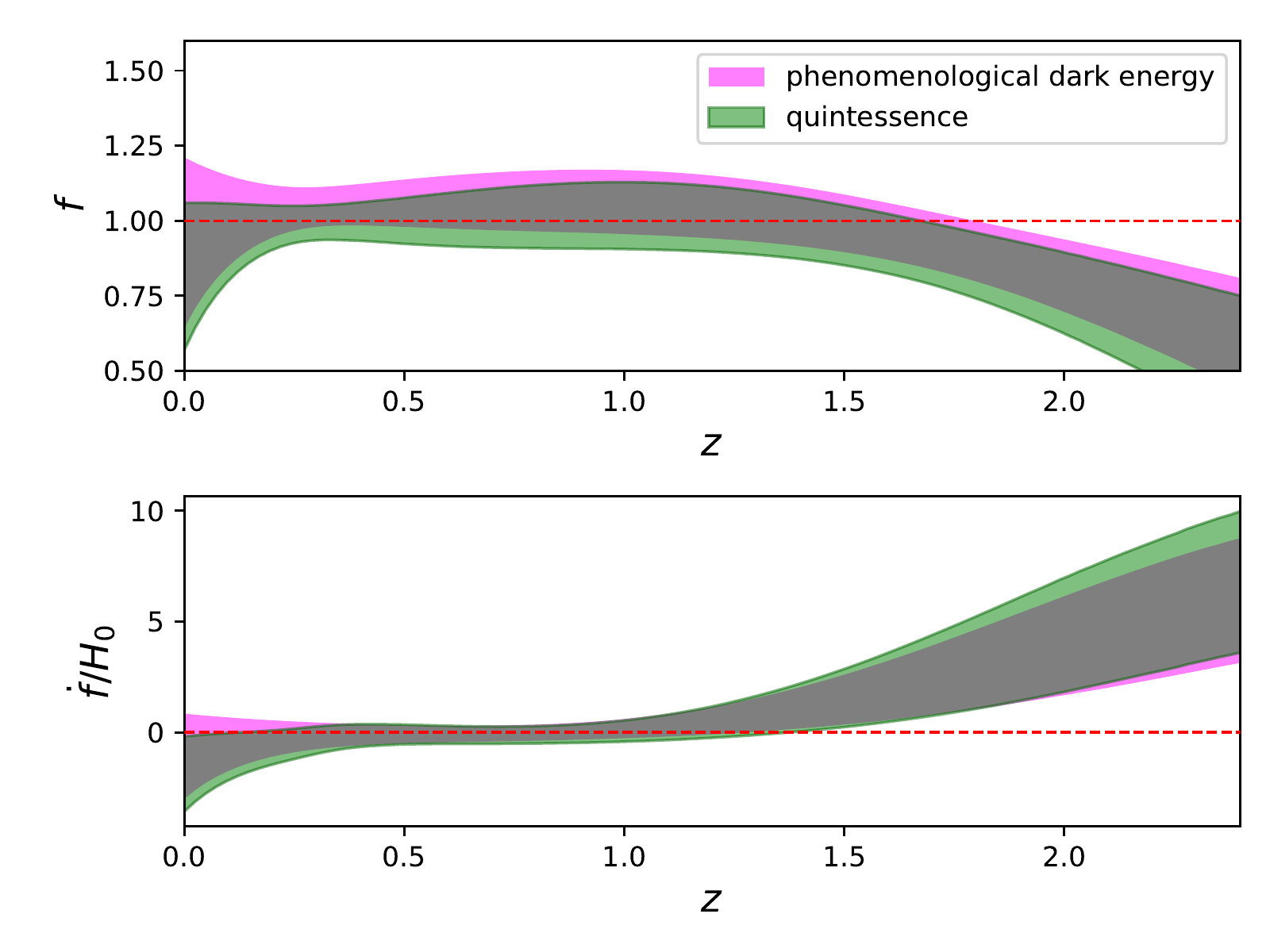} 
\includegraphics[width=0.45\textwidth]{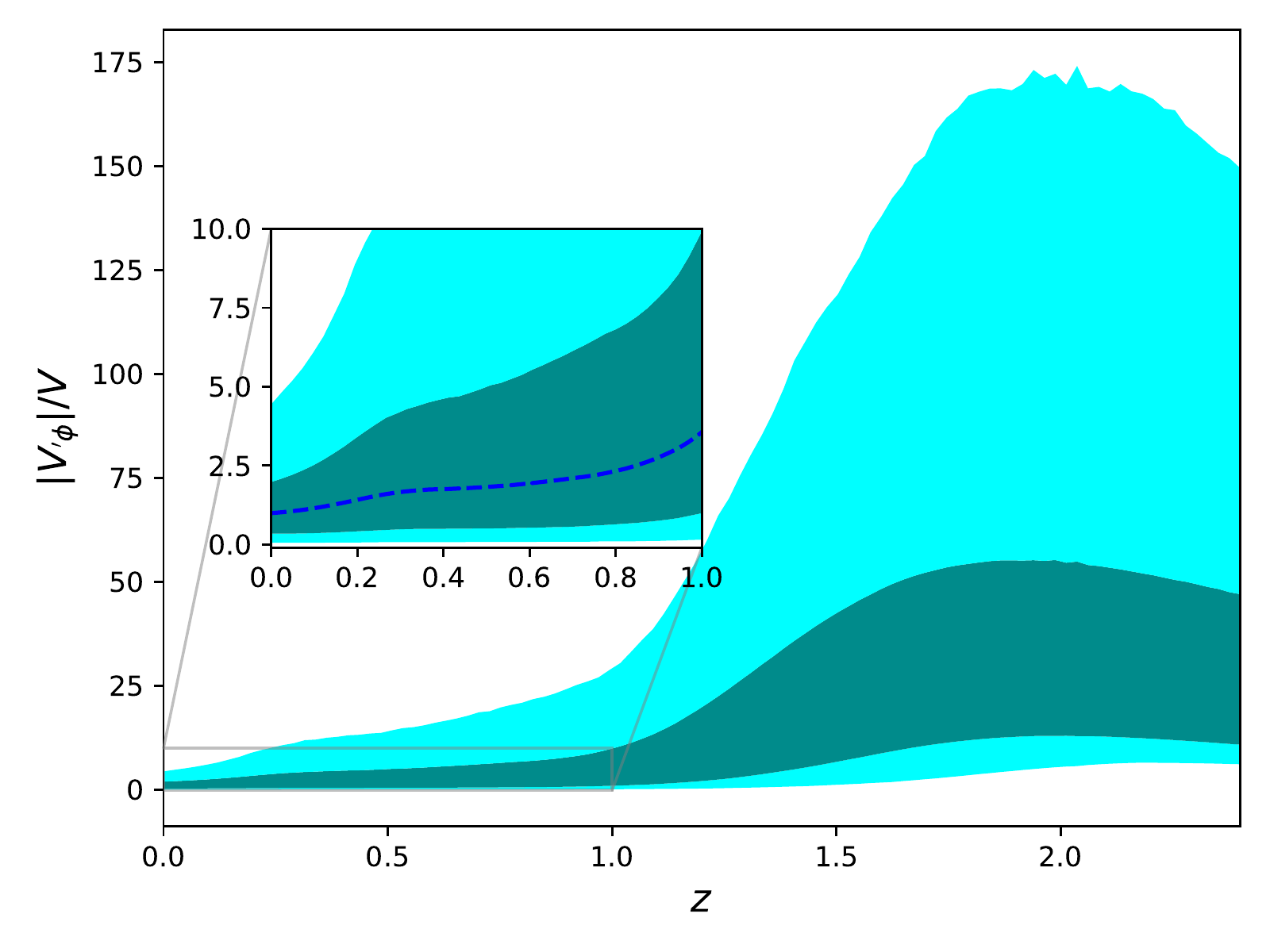}
\caption{Reconstructions of the dark sector interactions (left) and Swampland criterion $C1$ (right) for the coupled dark energy. We only show $68\%$ confidence level for the dark sector interactions. The red dashed line denotes a null interaction. The priors of $w(z)=w_0+w_a\frac{z}{1+z}$ are extracted from Planck. }
\label{fig:coupled}
\end{figure*}

%
%
%
For coupled dark energy, as we demonstrate in Sec.~\ref{sec:method}, the priors of either the equation of state of dark energy or the dark sector interactions should be given to reconstruct the cosmological parameters. Since there are various forms for the small (generally speaking) couplings, setting priors on coupling is not rational.  In this paper we use the priors of the dynamic equation of state of dark energy which has definite parametrizations such as Chevallier-Polarski-Linder (CPL) form and has been constrained or reconstructed from different observations~\citep{Aghanim:2018eyx,Ade:2015rim,Zhao:2017cud,Sahni:2014ooa}.   We extract the constraints of CPL form $w(z)=w_0+w_a\frac{z}{1+z}$ from Planck~\citep{Aghanim:2018eyx} with the corresponding released covariance of $w_0$ and $w_a$~\footnote{\url{https://pla.esac.esa.int}}. 
The reconstructions of the dark sector interactions and Swampland criteria $C1$ are shown in Fig.~\ref{fig:coupled}. Our results suggest a $1 \sigma$ signal of non-zero dark sector interactions for the coupled quintessence at low redshift and a $2 \sigma$ signal at $z>2$ for both the phenomenological dark energy and quintessence. In this paper we do not assume any cosmological model for either the dark energy or the couplings. Traditionally the interaction term is a parametric form of $Q$ in the literature, where $\dot \rho_c+3H\rho_c=Q\sim \dot f$. Though we plot the interaction results for both $f$ and $\dot f$ , $f$ is less meaningful since $f$ is only defined up to a constant that can be absorbed in $\Omega_c$.  Note that whether $\dot f$ is zero or not is independent of the values of the the density parameters. The upper bound of the Swampland criterion $C1$ at $z=0$ is 4.44 at $95\%$ significance. 

\section{Conclusions and Discussions}

In this paper, we investigated a model-independent approach to reconstruct the coupled dark energy and Swampland criteria in the context of the expansion of the Universe. We showed how the dynamics of dark energy and the dark sector interactions are degenerated and counterbalance each other in a fixed Universe expansion. Based on the expansion data from cosmic chronometers, BAO and SNe Ia,  our results suggest a $1\sigma$ and $2\sigma$ non-zero dark sector interaction at low ($z<0.3$) and high ($z>2$) redshift, respectively, for the coupled quintessence.  The Swampland criteria cannot be ruled out at present stage.

The mild deviations of a nonzero $\dot f $ that emerge at low redshift come from the compensation of the bias we stick to quintessence $w>-1$. This can be regarded as an indication of the preference of $\Lambda$CDM over quintessence. The inconsistency between Ly$\alpha$ and $\Lambda$CDM model at high redshift have been alleviated from $2.3 \sigma$ to $1.7 \sigma$ significance~\citep{Aghanim:2018eyx} with the latest update from eBOSS DR14~\citep{Agathe:2019vsu,Blomqvist:2019rah}. In this paper we can interpret the inconsistency as a signal of dark sector interactions. 

The flow of the dark sector interaction from dark matter to dark energy in the quintessence model at the low redshift is consistent with other model-fitting results~\citep{Costa:2016tpb,DiValentino:2019ffd}. As we said this is a result of the restriction for $w>-1$ (see also~\citep{Yang:2018euj}). While the flow from dark energy to dark matter induced by Ly$\alpha$ at high redshift is caused by the anomalous lower value of $H(z)$ for Ly$\alpha$ at $z=2.34$. This contradicts the interacting quintessence proposed by~\citet{Agrawal:2019dlm} from string theory to alleviate the Hubble tension. All implications of our results cast a shadow over quintessence model.

Assuming no dark sector interaction, reconstructions of the dynamics of dark energy are consistent with the cosmological constant. While setting priors on the equation of state from Planck, which is consistent with $w=-1$, Ly$\alpha$ BAO at $z=2.34$ gives a $2\sigma$ deviation from the null dark sector interaction at $z>2$. This means the dark sector interaction is more sensitive to the data anomaly than the dynamics of dark energy in our approach. This can be seen from the Eq.~(\ref{eq:dotf}), where the interactions rely on the second derivative of the Hubble parameter. This inspires that the evidence of the dynamics of the dark energy presented by~\citet{Zhao:2017cud} can also be interpreted as a signal of null-zero dark sector interactions.

One may note the choice of the priors of $w(z)$ from Planck is actually derived from the uncoupled dark energy model. At first glance this assumption may not be general and would influence the robustness of our analysis. However, in this paper we would like to reconstruct the dark sector interaction and swampland criteria for any reliable and reasonable knowledge of equation of state of dark energy we believe. Moreover, considering the coupled dark energy like~\citep{Costa:2016tpb}, the constraints of $w$ are pretty consistent with Planck. Actually, we can see the inclusion of dark sector interaction would not change the constraints of equation of state significantly (as we have said above the interaction is more sensitive than the equation of state). Thus we can conclude that our choice of $w(z)$ from Planck would not influence the robustness of our analysis in this paper. Our methodology could adopt any new priors of $w(z)$ constrained from the coupled or uncoupled dark energy in the future. 

Interestingly, from the model-independent reconstruction the anomalous Ly$\alpha$ BAO at $z=2.34$ gives a lower value of $H_0=67.72~\rm km~s^{-1}~Mpc^{-1}$ which is very consistent with Planck. While in the case without Ly$\alpha$, $H_0=69.03~\rm km~s^{-1}~Mpc^{-1}$ is consistent with the local SNe Ia calibrated by Tip of the Red Giant Branch (TRGB)~\citep{Freedman:2019jwv,Freedman:2020dne}. However, our approach is insensitive to the value of $H_0$ which is cancelled in the reconstructions of $f$, $c$ or $\tilde c$. As for $\dot f/H_0$, the rescaling will not influence the statistics of the deviations from 0.

Our methodology in this paper is very general in the context of the model-independent approach. In principle, any machine-learning techniques applied on the regression or reconstruction of functions from the large data sets are suitable for our approach. The Deep Learning~\citep{George:2017pmj,Escamilla-Rivera:2020fxq} and Artificial Neural Networks~\citep{Wang:2019vxv} developed recently provide rich alternatives of applications in cosmology.

We have mentioned that distance data can also be incorporated in the formulae. Actually from Gaussian process, the Hubble parameter, which is inversely proportional to the derivative of distance, can be combined with the luminosity distance and angular diameter distance data sets to jointly reconstruct cosmological parameters. This gives a possibility of using standard candles such as quasars~\citep{Risaliti:2018reu} and gamma ray bursts~\citep{Demianski:2016zxi} at high redshift. In addition, future cosmic probes such as Euclid~\citep{Amendola:2016saw}, DESI~\citep{Levi:2019ggs} and gravitational wave standard sirens~\citep{Abbott:2017xzu,Cai:2016sby,Cai:2017yww} can play very import role in the test of $\Lambda$CDM model at very high redshift.

\begin{acknowledgements}
TY thanks Jia-Jun Zhang, Yun-Long Zhang and Eoin \'O Colg\'ain for helpful discussions. We also thank the referee for improving the quality of this paper. This work is supported by an appointment to the YST Program at the APCTP through the Science and Technology Promotion Fund and Lottery Fund of the Korean Government, and the Korean Local Governments - Gyeongsangbuk-do Province and Pohang City.
\end{acknowledgements}

\bibliography{ref}

\end{document}